\newcommand{\simlt}{\lower.5ex\hbox{$\; \buildrel < \over \sim \;$}}
\newcommand{\simgt}{\lower.5ex\hbox{$\; \buildrel > \over \sim \;$}}

\documentclass{iaus}
\usepackage{graphicx}

\title[Luminous and Dark Matter in Galaxies] 
{Mapping the distribution of luminous and dark matter in strong lensing
galaxies}

\author[Ferreras \etal\ ]   
{Ignacio Ferreras$^1$%
  \thanks{email: ferreras@star.ucl.ac.uk},
 Prasenjit Saha$^2$, Liliya L. R. Williams$^3$\break  \and Scott Burles$^4$}

\affiliation{$^1$Dept. of Physics, King's College London, Strand, 
London WC2R 2LS, UK\\
[\affilskip]$^2$Institute for Theoretical Physics, University of Z\"urich, 
Winterthurerstr. 190, CH8057 Z\"urich, Switzerland\\
[\affilskip]$^3$Dept. of Physics, University of Minnesota, 116 Church St. SE,
Minneapolis, MN55455, USA\\
[\affilskip]$^4$Dept. of Physics, Massachusetts Institute of Technology,
77 Massachusetts Ave., Cambridge, MA02139, USA}

\pubyear{2007}
\volume{244}  
\pagerange{1--10}
\date{?? and in revised form ??}
\jname{Dark Galaxies and Lost Baryons}
\editors{J. Davies et al., eds.}
\begin{document}

\maketitle

\begin{abstract}
We present the distribution of luminous and dark matter in a set of
strong lensing (early-type) galaxies. By combining two independent
techniques -- stellar population synthesis and gravitational lensing
-- we can compare the baryonic and dark matter content in these
galaxies within the regions that can be probed using the images of the
lensed background source. Two samples were studied, extracted from the
CASTLES and SLACS surveys. The former probes a wider range of
redshifts and allows us to explore the mass distribution out to $\sim
5R_e$. The high resolution optical images of the latter (using
HST/ACS) are used to show a pixellated map of the ratio between total
and baryonic matter. We find dark matter to be absent in the cores of
these galaxies, with an increasing contribution at projected radii
$R\simgt R_e$. The slopes are roughly compatible with an isothermal slope
(better interpreted as an adiabatically contracted NFW profile), but a
large scatter in the slope exists among galaxies. There is a trend
suggesting most massive galaxies have a higher content of dark
matter in the regions probed by this analysis.

\end{abstract}

\firstsection 
\section{Introduction}

The standard paradigm of galaxy formation rests on the concept of
the dark matter halo as the building block of structure. This dark matter
halo dominates the mass content -- with the cosmological ratio between 
baryonic and dark matter around $\sim$1/6 
(\cite[Spergel \etal\ 2006]{WMAP}) -- although this primordial ratio will change
because of the ``baryon'' physics that can drive infall or outflows of
baryons towards/away from the central regions.

The presence of dark matter on galaxy scales is best shown by the
rotation curves of disk galaxies. These galaxies feature a coherent
rotational motion (i.e. dynamically ``cold'' systems) with a prominent flat
portion in the outer regions away from where most of the baryons live
(see e.g. \cite[Sofue \& Rubin 2001]{Sof01}).

On the other hand, the halos of elliptical galaxies are harder to
study as the dynamics of these systems is ``hot'', i.e. mainly
supported by velocity dispersion. The dynamics of
elliptical galaxies suggests a formation process from major mergers
which erases the coherent motion from disk-like progenitors,
creating a smooth, concentrated distribution of stars supported by
pressure. In a cold dark matter scenario, the most massive elliptical
galaxies should assemble at late times, in stark contrast with
the overall coeval,  homogeneously old and metal rich stellar populations found
throughout early-type galaxies (see e.g. \cite[Trager \etal\ 2000]{SCT00}). 
Furthermore, $\alpha$ element-to-Fe abundance ratio studies
suggest the high metallicity stellar populations are not only formed 
at early times but over a very short period, comparable with the dynamical
timescale for the collapse of a mass corresponding to a giant elliptical
galaxy. Simple back-of-the-envelope estimates applied to recent datasets
(\cite[Lintott \etal\ 2006]{LFL06}) still support the old idea of 
early collapse of \cite{Lar75}.

Detailed kinematic analyses such as those from the SAURON
collaboration have shown interesting features such as kinematically
decoupled cores -- which would suggest a recent merger -- in galaxies
otherwise populated by an old and very homogeneous stellar population
-- which would suggest no star formation (\cite[Davies \etal\ 2001]{NGC4365}).
The complexity of the
``baryon physics'' controlling galaxy formation prevents us from
getting an accurate picture using {\sl ab initio} modelling. It is
therefore of paramount importance to understand the structure of the halos
of early-type galaxies. In this paper we present ongoing research on
the resolved structure of the dark matter halos of early-type systems
using a combination of gravitational lensing and stellar population synthesis.

\section{Measuring total masses in elliptical galaxies}\label{sec:totmass}

\subsection{Stellar kinematics} 
The spectral absorption features of the stellar populations can be
used to determine the line-of-sight velocity dispersion ($\sigma$).
The first-order result one can get is a luminosity-weighted 
measure ($\langle\sigma\rangle_e$) which leads to a rough estimate 
of the total mass, namely: $M=5\langle\sigma\rangle_e^2R_e/G$ that 
happens to give a good approximation to more sophisticated 
modelling (\cite[Cappellari \etal\ 2006]{Cap06}).

A proper comparison with dynamical models require knowledge of the
radial profile $\sigma (R)$. The observations are compared with models
that parametrise the galaxy as a halo density distribution (commonly a
NFW profile) and a baryonic density distribution (a Hernquist profile
for early-type galaxies). This distribution is used to solve for the
profile of the velocity dispersion using Jeans' equation, with
assumptions made about the anisotropy of the velocity dispersion
tensor -- see \cite{Mam05} for a detailed analysis.

\subsection{Planetary Nebul\ae}
Planetary nebulae (PNe) in the outer regions of galaxies can be used
as dynamical tracers of the gravitational potential well. Given that
PNe are just evolved stars, one can consider them as unbiased
tracers. The characteristic emission lines of PNe can be used to
select these stars in the outskirts of galaxies.  The kinematics can
be obtained directly from the Doppler shifted lines. The Planetary
Nebula Spectrograph (PN.S; \cite{PN.S}) has exploited this feature to
target several nearby early-type galaxies. The first results
(\cite[Romanowsky \etal\ 2003]{Rom03}) have revealed decreasing 
velocity dispersion profiles. The inferred mass-to-light ratios out to
$5R_e$ are low enough to be explained only by the stellar mass content.

This claim has been challenged by \cite{Dek05}, who claim that the PNe 
observed in the outskirts are biased tracers of the halo and preferentially
lie on orbits with higher anisotropy. However, non-isotropic orbits
can still be accommodated within a model with little dark matter
within $5R_e$ (e.g. \cite[Douglas \etal\ 2007]{Doug07})
In this symposium, the contribution by 
Napolitano has presented the latest work done by the PN.S team.

\subsection{X-rays}
Elliptical galaxies are surrounded by a tenuous hot gas which makes
the galaxy bright in X-rays from bremsstrahlung. This has been
traditionally the observable used to probe the mass in elliptical
galaxies out to distances well away from the optical size. X-ray
observations give the projected radial profile in density and
temperature of this hot gas. Hydrostatic equilibrium and spherical
symmetry are invoked in order to determine the total mass of the
galaxy. This method is of course biased towards galaxies bright enough
in X-rays to have an accurate measure of the radial profiles. Masses
determined this way result in mass-to-light ratios $M/L_B>40$ (see
e.g. \cite[O'Sullivan \& Ponman 2004]{OP04}; \cite[Loewenstein \&
Mushotzky 2003]{Loew03}) which require a dominant contribution from
dark matter.

The X-ray measurements are in remarkable contrast with the PNe
estimates discussed above -- which are extended out to similar 
distances, i.e. $R\simlt 5R_e$. 
Systematic effects could be related either to the method
employed in translating the observations into actual masses, or to the
way the sample is selected. For the method based on PNe, one needs to
select targets close enough for the PN.S instrument.  On the other
hand, X-ray measurements must target X-ray bright galaxies, which will
bias the sample towards massive systems. 

\subsection{Globular Clusters}
Another method currently used to determine the masses of early-type
galaxies consists of a kinematic analysis of their globular cluster
system. This method is complementary to the PNe studies but one must
consider whether globular clusters are indeed unbiased tracers of the
gravitational potential (cf. \cite[Prieto \& Gnedin 2006]{PG06}). 

Recent estimates of the mass of NGC4636 (\cite[Schuberth \etal\ 2006]{Schu06}) 
give values of $M/L_B\sim 20$ compared to a factor of two higher values 
using X-rays (\cite[Loewenstein \& Mushotzky 2003]{Loew03}). 
However, from galaxy to galaxy the radial profile of velocity dispersion 
of GCs show a wide scatter, with some galaxies presenting a decreasing trend
with radius, analogous to PNe studies (\cite[Romanowsky 2006]{Rom06}).

\subsection{Strong gravitational lensing}
 
Gravitational lensing relies on a very different physical effect from
the above, which is the deflection of light from a background source
by $\sim GM/c^2R$ radians, where $M$ is the projected mass enclosed
within projected radius $R$.  By a happy coincidence, for galaxies at
$z\simgt 0.1$, the deflection is about right to produce
multiple images of sources that are directly behind.  Notice that
while strong lensing depends on projected quantities, the deflection
angle (and hence the image separations) scale in the same way as the
velocity variance.  As a result lensing as a mass tracer is not
dissimilar to stellar kinematics.

There is a large literature on detailed modelling of lenses. One
strategy is to fit to a parametrized model based on stellar
dynamics. This technique goes back to the very first lens-modelling
paper (\cite[Young \etal\ 1981]{Yo81}) but in recent years has seen
progressively more complex parametric forms
(\cite[e.g., Bernstein \& Fischer 1999]{BF99}). The most commonly used
parametric model families are described in \cite{CK01},
with fitting implemented in the associated GRAVLENS software.  Other
techniques are multipole expansions (\cite[Trotter \etal\ 2000]{Trot00}) or
free-form -- non-parametric -- models (\cite[Saha \& Williams 1997]{SW97}). The
latter have seen increased popularity in recent years
(e.g. \cite[Diego \etal\ 2005]{JM05}; \cite[Koopmans 2005]{Ko05};
\cite[Suyu \& Blandford 2006]{SB06}).

\begin{figure}
 \centerline{
   \scalebox{0.4}{\includegraphics{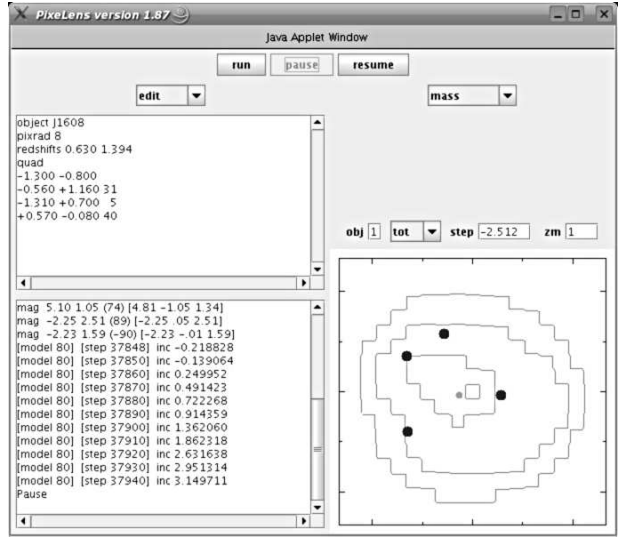}}
 }
 \caption{PixeLens is a Java-based application which can
be run over the web ({\tt http://www.qgd.unizh.ch/programs/pixelens/}).
One of the lenses in our sample is shown. The box on the top-left
side of the window includes the information about the lens
(ID, pixel size, redshifts, image positions and time delays). 
After 80 models are run, the figure on the right shows contours
of mass density along with the image positions and the centre
of the lensing galaxy in grey.}
 \label{fig:PixLens}
\end{figure}

The main difficulty in lens modelling is that the models are highly
non-unique.  The enclosed mass within the region of the images is
usually very well-constrained, and likewise the direction of the
ellipticity, but the steepness of the radial profile is extremely
model-dependent. This was first noted by \cite{Koch91},
and has been increasingly reinforced by later modelling work, and why
this kind of model-dependence should arise can be understood by
considering qualitative aspects of lensing theory
(\cite[Saha \& Williams 2003]{SW03}).  The problem of model degeneracies has
led to techniques involving large ensembles of models to explore those
degeneracies.  Of these techniques, the method of pixelated model
ensembles introduced by \cite{WS00} and subsequently
made available in the PixeLens software (\cite[Saha \& Williams 2004]{SW04}) is
the one used in the present work.  Similar ideas using parametric
models have also been developed (\cite[Keeton \& Winn 2003]{KW03}).

\section{Non-parametric lensing estimates}\label{sec:lensmass}
Our analysis of the surface mass density is done using 
a well-developed and tested 
non-parametric method that uses the image positions of the
background source and their time delays -- whether available --
to generate an ensemble of pixellated mass maps of the lensing
galaxy. A few -- very mild -- constraints are further imposed on
the generated models, namely:
\begin{itemize}
\item The gradient of the surface mass density should 
  point within 45$^o$ towards the centre.
\item No pixel shall have more than twice the surface density
  of its neighbours (except for the central pixel, which is unconstrained).
\item The radial density gradient must be steeper than $R^{-0.5}$.
\end{itemize}
The ensemble is used to determine the probability distribution
function of the pixellated surface mass density. We refer the
reader to \cite{SW04} for a detailed analysis
of this method. We emphasize that only the positions and time
delays are used to constrain the surface mass density. 
We assume the background
source to be a point-like object and we do not explore the issue
of flux ratios between the images. 

This method has been written as a Java-based application -- PixeLens --
with a user friendly interface. PixeLens can be run as an applet
from a web browser (see figure~\ref{fig:PixLens}).

\section{Measuring stellar masses}\label{sec:starmass}
\begin{figure}
 \centerline{
   \scalebox{0.4}{\includegraphics{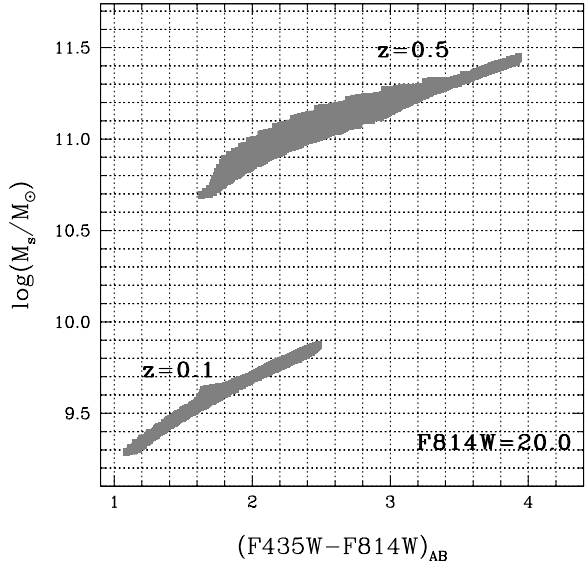}}
 }
 \caption{This figure illustrates the estimate of the stellar mass
content from photometry. A grid of $\tau$-models (exponentially
decaying star formation history) are generated from the population
synthesis models of \cite{BC03}. The grey regions encompass all
possible values of $B-i$ colour vs. stellar mass for a F814W=20 galaxy
at redshift z=0.1 (bottom region) or z=0.5 (top region). Even though the
degeneracies between stellar population parameters are very large with
broadband photometry, one can get stellar masses within 0.2-0.3 dex.}
\label{fig:Mstar}
\end{figure}

Stellar masses are calculated from the (optical and/or NIR) photometry of the
lenses. In both samples high resolution followup imaging is available
from the Hubble Space Telescope. The lenses were observed in at least 
two filters, which allows us to determine the mass-to-light ratio of
the underlying stellar populations. Figure~\ref{fig:Mstar} shows our
methodology for stellar mass estimates. The grey areas span
the allowed regions in a colour-stellar mass diagram for a wide
range of $\tau$ models (exponentially decaying star formation history).
The composite populations are obtained from the stellar population 
synthesis models of \cite{BC03}, taking a \cite{chab03} IMF. A wide range of
ages and metallicities are considered (see \cite[Ferreras \etal\ 2005]{FSW05} 
for details). 
Even though the inherent degeneracies with broadband photometry
prevent us from extracting ages and metallicities, we can nevertheless
determine the M/L ratio with acceptable uncertainty. Figure~\ref{fig:Mstar} 
corresponds to a F814W=20 galaxy at redshifts z=0.1 (bottom region) or
z=0.5. The stellar masses are determined with a 
$\Delta\log (M_{\star}/M_\odot)\sim$0.2--0.3 dex uncertainty.

\section{The sample}\label{sec:sample}
The lensing galaxies presented here were extracted from 
CASTLES\footnote{CASTLES webpage: {\tt http://cfa-www.harvard.edu/castles}}
(CfA-Arizona Space Telescope Lens Survey; \cite[Rusin \etal\ 2003]{Ru03}) and
from SLACS\footnote{SLACS webpage: {\tt http://www.slacs.org}}
(Sloan Lens ACs Survey; \cite[Bolton \etal\ 2006]{bo06}). The former targeted a
sample of known lensing system with the Hubble Space Telescope
both in the optical and NIR. SLACS used the Sloan Digital Sky Survey
to systematically look for nebular emission lines at higher redshift than the
targeted galaxies, which were selected based on the redshift dependence
of the cross section to strong lensing. The candidates were subsequently
imaged with HST/ACS. The CASTLES dataset spans a wider range of redshifts
and the lensing galaxies can be explored out to larger radii, in some
cases out to $5R_e$. Our analysis of a subset of 18 CASTLES lenses
can be found in \cite{FSW05} whereas the detailed analysis of 9 SLACS
will be published soon (Ferreras, Saha \& Burles, submitted).

\section{Dark Matter vs. Baryonic Matter in elliptical galaxies}\label{sec:profs}

For the sample of lensing galaxies described above we compute the
total surface mass density with PixeLens using a reasonable pixel size
(typically $21\times 21$ pixels for the region over which masses can be
reliably determined). The optical and NIR (if available) images are
then used to determine the surface brightness and colors within a given
PixeLens pixel. The colours are compared with a large grid of $\tau$
models as explained above, spanning a range of formation epochs
($1<z_{\rm FOR}<5$), star formation timescales
($-1<\log\tau($Gyr$)<+1$) and metallicities ($-1<\log Z/Z_\odot
<+0.3$). We refer the interested reader to \cite{FSW05} and references
therein for details of the process. The total and stellar mass maps
are subsequently registered and compared.

\begin{figure}
 \centerline{ \scalebox{0.8}{\includegraphics{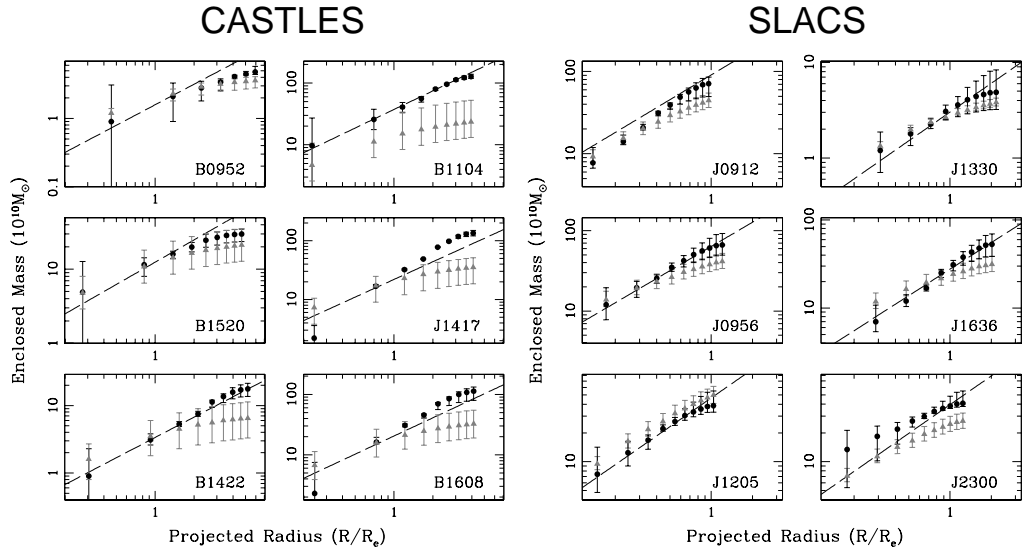}} }
 \caption{Radially-averaged cumulative 2D mass profiles for a set
of galaxies from the CASTLES (left) and the SLACS (right) samples.
The grey triangles (black dots) represent the stellar (total) mass content.
Error bars are shown at the 90\% confidence level. The dashed line
corresponds to the projection of a spherically symmetric distribution with 
density $\rho (r)\propto 1/r^2$. Notice the CASTLES
galaxies are separated into low-mass (left panels) and high-mass
(right panels).}
 \label{fig:profs}
\end{figure}

Figure~\ref{fig:profs} shows the radially averaged cumulative profiles
in stellar (grey triangles) and total mass (black dots). The error bars
are shown at the 90\% confidence level. The radial distance is expressed
as a ratio with the effective radius. The dashed line in all panels
correspond to an isothermal profile, i.e. the projection of a 
spherically symmetric distribution with density $\rho (r)\propto 1/r^2$.
The slopes -- which can be interpreted as an adiabatically contracted
NFW profile from the effect of the collapse of the baryons -- are
in agreement with previous work using parametric methods (e.g.
\cite[Rusin \etal\ 2003]{Ru03};\cite[Koopmans \etal\ 2006]{Ko06}).
Our method takes the uncertainties fully into account, showing that
such lens systems cannot be used to robustly infer the slope of 
the density profile.
Having sources at multiple redshifts can break the degeneracies and
constrain the profiles extremely well (\cite[Saha \etal\ 2006]{PS06}),
but this is a luxury available in a very few rich clusters, and no
known galaxies so far.

The CASTLES galaxies shown in the figure are subdivided into low-mass
(leftmost panels) and high-mass systems. Notice the significant difference
between the profiles in these two subsets. The dark matter contribution
in the halos of massive galaxies is larger than in lower mass galaxies. 
This point will be discussed below.

\begin{figure}
 \centerline{
   \scalebox{0.4}{\includegraphics{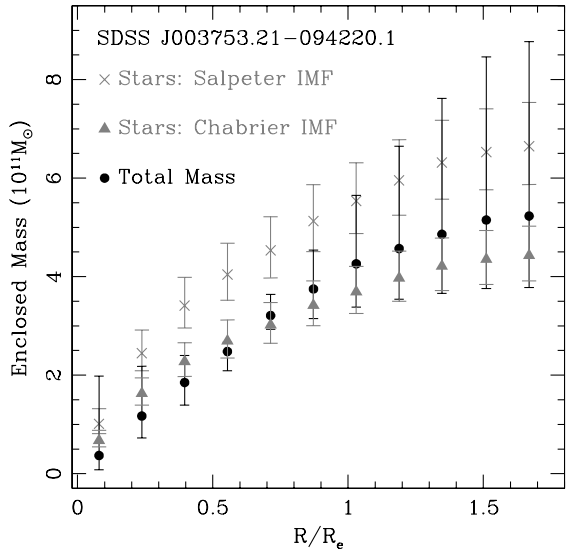}}
 }
 \caption{Radially-averaged profile of lens J0037-09 from the SLACS
sample.  We compare the stellar mass estimates from our fiducial
Chabrier initial mass function (IMF; grey triangles) and the standard
power law of Salpeter (grey crosses). Even though the high-mass end of
the Salpeter IMF is compatible with many observations, its
extrapolation as a single power law down to the H core burning limit
($\sim 0.1M_\odot$) gives values of the stellar masses which cannot be
accounted for by the lensing data (black dots).}
 \label{fig:prof1}
\end{figure}

Figure~\ref{fig:prof1} compares the stellar masses predicted by a
different initial mass function (IMF). By default, we use the
\cite{chab03} IMF which is a realistic approximation both at the low-
and the high-mass ends. A \cite{salp55} IMF describes the mass
function by a single power law across the full 0.1-100$M_\odot$ stellar
mass range.  This approximation has been widely used in studies of
stellar populations.  A single power law is a valid approximation when
exploring stars at the high-mass end. Low mass stars contribute
very little to the total luminosity of the population but they can
contribute very significantly to the total stellar mass when assuming
an extrapolation of the Salpeter power law down to the threshold of hydrogen
core burning ($\sim 0.1 M_\odot$). Hence, photo-spectroscopic data
from unresolved populations cannot reliably constrain the slope at low masses. 

Observations of the local stellar census suggest a turnoff from a power
law at masses above $1M_\odot$ to a lognormal distribution below
(\cite[Chabrier 2003]{chab03}). Furthermore, theoretical
arguments (\cite[Larson 2006]{Lar06}) also hint at a characteristic
mass scale in the IMF which depends on the physics of the collapse.
At the high mass end, a power law can be justified if massive stars
form by scale-free accretion in dense environments.
Our comparison of stellar and total mass rules out a Salpeter IMF 
as illustrated in figure~\ref{fig:prof1}. This result is in agreement with
recent work by the SAURON group (\cite[Cappellari \etal\ 2006]{Cap06}). 
We emphasize that the total mass (lensing) and the stellar mass
content are calculated from independent observables. The former ONLY
uses the image positions and time delays and the latter uses the
photometry of the lensing galaxy. The fact that we can tackle the
issue of the stellar IMF illustrates that the systematic uncertainties
of the combination of lensing and stellar mass estimates must be
small.

\begin{figure}
 \centerline{
   \scalebox{0.7}{\includegraphics{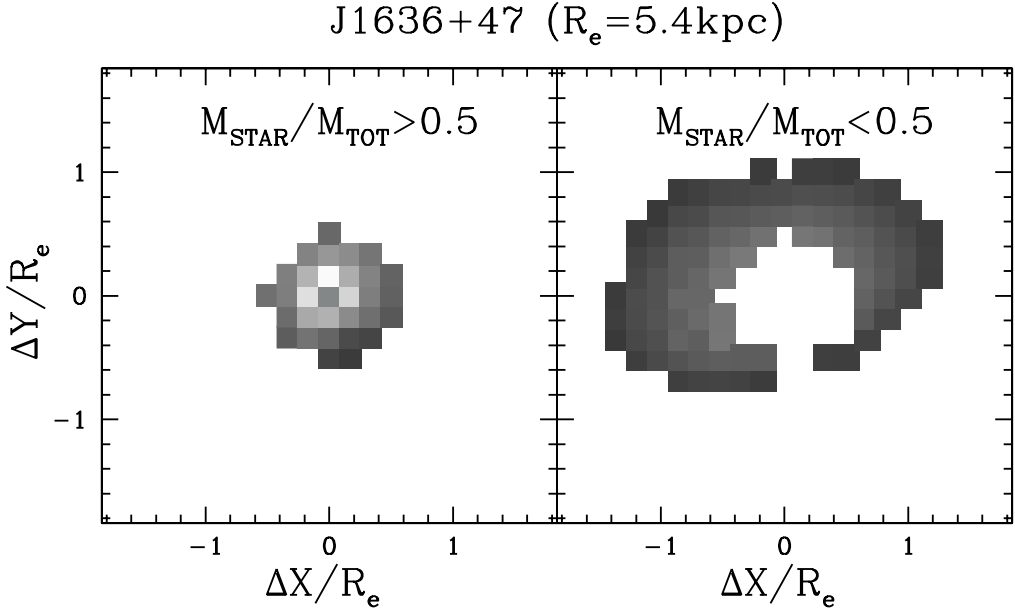}}
 }
 \caption{Mass map of lens J1636+47 from SLACS. The greyscale 
corresponds to the surface density in total (lensing) mass.
The panels separate regions corresponding to the ratio between
baryon and total mass, so that the panel on the left shows the
region where the mass budget is dominated by baryons.}
 \label{fig:map}
\end{figure}

Our lensing analysis of the SLACS galaxies do not extend further 
than $R\simlt 1$--$2R_e$. However, the lower redshift of these
galaxies allows us to perform a detailed 2D map of the contribution of
dark matter. In an upcoming paper (Ferreras, Saha \& Burles,
submitted) we present false colour maps of the nine galaxies studied
from SLACS.  Figure~\ref{fig:map} shows in greyscale the mass map of
lens J1636+47 (z=0.23). The greyscale corresponds to the total surface
mass density, and the projected scale is given as a fraction of the
effective radius ($R_e=5.4$kpc).  We divide the map into two
regions. The panel on the left shows the region where $M_{\rm
STAR}/M_{\rm TOT}>0.5$ i.e. mostly dominated by baryons, whereas the
panel on the right shows the region dominated by dark matter. As
expected, the core is dominated by baryons -- as seen in the radial
profiles of figure~\ref{fig:profs} -- and the transition takes place
at around $R_e$. In Ferreras, Saha \& Burles we perform an analysis of
the mass distribution of the SLACS lenses and find that in all galaxies
the ellipticity of the baryonic and total surface mass density 
distributions do not differ significantly.

\section{Conclusions}\label{sec:end}

Strong gravitational lensing works as a very powerful -- albeit restricted --
``dark matter'' telescope which can give us very valuable clues about the
process of galaxy formation within halos. Unfortunately one cannot
apply for ``dark matter telescope time'' to observe a specific sample,
and one can only take what Nature gives: galaxies that happen to
lie along the line of sight towards bright background sources. Nevertheless,
the growing number of strong lenses observed in detail by HST has allowed us to
determine in some detail the dark matter distribution in massive
early-type galaxies.

\begin{figure}
 \centerline{
   \scalebox{0.8}{\includegraphics{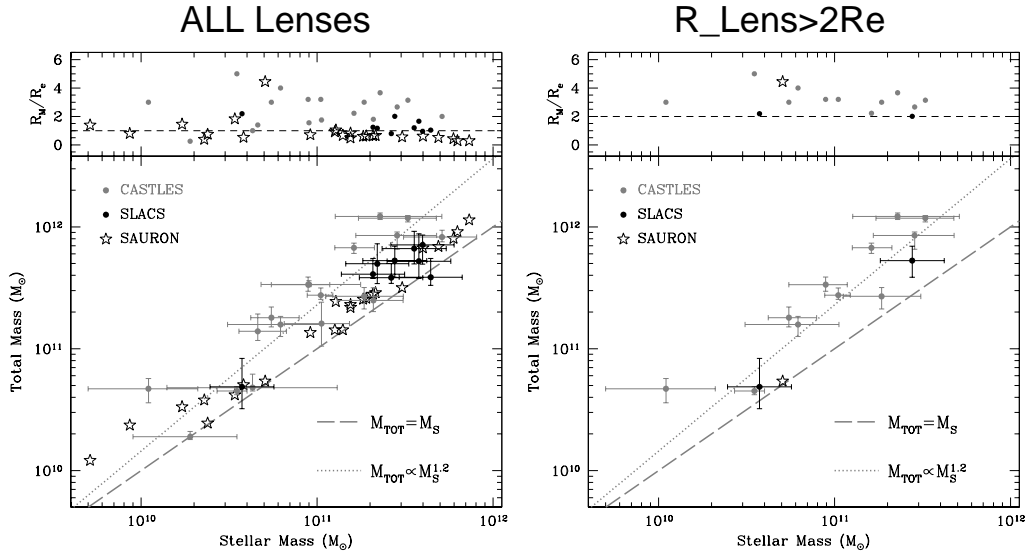}}
 }
 \caption{Comparison of the total and stellar mass enclosed within R$_M$.
 Our samples are shown along with the detailed dynamical analysis of 
 \cite[Cappellari \etal\ (2006)]{Cap06} using SAURON data. The complete
 sample is shown on the left. The dashed line follows the 1:1 correspondence 
 between total and stellar mass, whereas the dotted line -- which is not
 a fit -- is a power law $M_{\rm TOT}\propto M_{\rm STAR}^{1.2}$ which 
 explains the tilt of the Fundamental Plane (\cite[Ferreras \& Silk 2000]{FS00}).
 The sample is a ``mixed bag'' of systems measured at different radii
 (top panel). The figure on the right shows only those systems for which
 the observations probe further out than $2R_e$.
}
 \label{fig:mtot}
\end{figure}

Figure~\ref{fig:mtot} shows the stellar vs. total mass measured within
our lensing radius ($R_M$). This lensing radius is the extent over
which the surface mass density can be reliably measured using the
positions of the images of the background source. The top panel shows
$R_M$ with respect to the effective radius. 
Our CASTLES and SLACS samples are shown along with the recent
estimates from the SAURON group (\cite[Cappellari \etal\ 2006]{Cap06}).
The dashed line is a 1:1 mapping between total and stellar mass
(i.e. no dark matter required). The dashed line is NOT a fit, but the
trend that could explain the tilt of the Fundamental Plane 
(see e.g. \cite[Ferreras \& Silk 2000]{FS00}). This tilt cannot be 
accounted for simply by stellar populations and in our case a variation
of the dark matter contribution would be invoked. The panel on the left 
shows what a ``mixed bag'' the sample is. The SAURON data mostly
probe the inner regions, seldom extending out of $R_e$.
Similarly, our lower-redshift sample (from SLACS) also probes the
inner parts of galaxies. On the right hand panel, we only show those
galaxies for which the analysis extends at least out to $2R_e$. One
can see that the trend suggested by the dotted line could be a
reasonable fit to the data points (undoubtedly with a large
scatter). We conclude this paper by suggesting that this is a trend
that can explain the apparent mismatch between the mass estimates
gathered using either PNe or X-ray light. A selection of galaxies
using the latter technique will bias the sample in favour of massive
systems, which we claim to be more dominated by dark matter. On the
other hand, lower mass galaxies -- more likely to be observed by the PNe
studies which can only target nearby systems -- will have a smaller
contribution from dark matter in the inner regions. Recent work by
Napolitano and collaborators (see contribution in these proceedings)
targeting more massive galaxies with the PNe method seem to favour
this point.

%


\end{document}